\documentclass[prl,a4paper,aps,twocolumn,nofootinbib,nobibnotes,superscriptaddress,preprintnumbers]{revtex4}

\usepackage[dvips]{graphicx}
\usepackage{amsmath,amssymb,mathrsfs}
\usepackage{bm}
\usepackage{times}
\usepackage{epsfig}
\usepackage{verbatim}
\usepackage{bm}
\usepackage[utf8]{inputenc}
\usepackage{graphics}
\usepackage{graphicx,epsfig,amssymb,amsmath,color,cancel}
\usepackage{subfigure}
\usepackage[english]{babel}
\usepackage{soul}

\begin{document}


\title{High-Temperature Electroweak Symmetry Breaking by SM Twins}

\author{Oleksii Matsedonskyi}

\affiliation{Department of Particle Physics and Astrophysics, Weizmann Institute of Science, Rehovot 7610001, Israel}


\begin{abstract}

We analyse a possible adjustment of Twin Higgs models allowing to have broken electroweak (EW) symmetry at all temperatures below the sigma-model scale $\sim 1$TeV. The modification consists of increasing the Yukawa couplings of the twins of light SM fermions. The naturalness considerations then imply a presence of relatively light electroweak-charged fermions, which can be produced at the LHC, and decay into SM gauge and Higgs bosons and missing energy. Analysis of experimental bounds shows that such a modified model features an increased amount of fine-tuning compared to the original Twin Higgs models, but still less tuning than the usual pseudo-Nambu-Goldstone Higgs models not improved by $Z_2$ twin symmetry. The obtained modification in the evolution of the EW symmetry breaking strength can, in particular, have interesting implications for models of EW baryogenesis, which we comment on.

\end{abstract}

\maketitle

\section{Introduction}

The origin of the observed matter-antimatter asymmetry remains an open question of fundamental physics. Among various proposed scenarios to address this question, the electroweak baryogenesis~\cite{Shaposhnikov:1987tw,Cohen:1990it} (EWBG) stands out as the one unavoidably requiring sub-TeV-scale new physics beyond standard model (SM). This new physics is responsible for rendering the electroweak phase transition to be of the first order, and also for introducing new sources of CP violation. This prediction motivates numerous experimental searches (see e.g.~\cite{Curtin:2014jma}). 
However, in order to explore the idea of EWBG to the fullest, it is important to understand whether there are alternative realizations of the latter. In particular, the above prediction for the new physics scale is derived for models where the electroweak symmetry is restored at temperatures above around $160$~GeV. Such a symmetry restoration is driven by a positive correction to the Higgs mass induced by the interactions with the SM plasma
\begin{equation}\label{eq:SMSR}
\delta m_h^2(T)_{\text{SM}} \simeq T^2\left(\frac{\lambda_t^2}{4}  + \frac{\lambda_h}{2} + \frac{3 g^2}{16}+\frac{g^{\prime 2}}{16} \right) \simeq 0.4 \,T^2,
\end{equation}
with $\lambda_t, \lambda_h, g, g'$ being top quark Yukawa, Higgs quartic and EW couplings respectively.

Recently, several works~\cite{Meade:2018saz,Baldes:2018nel,Glioti:2018roy,Matsedonskyi:2020mlz} proposed scenarios where this is not the case, realizing high-temperature breaking of the electroweak symmetry\footnote{See~\cite{Weinberg:1974hy,Mohapatra:1979qt,Fujimoto:1984hr,Salomonson:1984rh,Salomonson:1984px,Dvali:1995cj,Bimonte:1995xs,Bimonte:1995sc,Dvali:1996zr,Orloff:1996yn,Pietroni:1996zj,Gavela:1998ux,Hamada:2016gux,Bajc:1998jr,Jansen:1998rj,Bimonte:1999tw,Pinto:1999pg} for other studies of symmetries broken at high temperature and in particular~\cite{Espinosa:2004pn,Ahriche:2010kh,DiLuzio:2019wsw,Aziz:2009hk} for the studies in Goldstone Higgs models.}, also referred to as symmetry non-restoration (SNR). With this mechanism in action, the transition from zero to large Higgs VEV can happen at much higher temperatures than was typically assumed, with the EW symmetry always being broken from that point till today. The new physics needed to be active during the transition to generate the first order phase transition, and produce CP violation, can be correspondingly heavier. The common prediction of the proposed models is a large number of new SM-neutral degrees of freedom which generate a negative temperature-dependent correction to the Higgs mass, thus counteracting the positive SM contribution of Eq.~(\ref{eq:SMSR}). These new states have to feature an appropriate type of couplings to the Higgs boson and a relatively low mass.  

Given the large multiplicity and very special features of the SNR-states, it is important to find whether there exist motivated theories which automatically predict their presence.  
The Twin Higgs models~\cite{Chacko:2005pe} contain this type of states by default: these are twins of SM quarks, leptons and gauge bosons. The thermal correction to the Higgs potential induced by the twin SM states is minimized at some large Higgs field value, in contrast to the correction induced by the SM states, which is minimized at the Higgs origin. In this note we derive the conditions needed for the twin contribution to dominate over the symmetry-restoring SM correction, thus producing SNR. One of the phenomenological predictions of the resulting model is a presence of EW-charged TeV-scale partners of the twin fermions. We discuss current and future experimental sensitivity to them. 

We dedicate a special attention to the fine-tuning of our SNR Twin Higgs model. The main motivation for considering the Twin Higgs scenarios, with a field content significantly enlarged with respect to the SM, is a possibility to decrease the amount of fine tuning, needed to generate the separation between the EW scale and the scale of SM-charged new physics, which is pushed up by the collider experiments. In this respect, the SNR Twin Higgs    
appears to be worse than the vanilla Twin Higgs scenarios but still better than the usual Goldstone Higgs models \cite{Panico:2015jxa} not enhanced by a $Z_2$ symmetry. While loosing in the tuning to the vanilla Twin Higgs, the model with new relatively light fermions allows for more options to relax the tensions with the electroweak precision tests~\cite{Grojean:2013qca} which are generic for Goldstone Higgs models.  

More generally, we provide for the first time a concrete quantitative evaluation of the fine tuning necessary to accommodate the EW symmetry non-restoration at high temperature, which was not given much of attention previously.

\section{High-Temperature EW symmetry breaking in Twin Higgs models}

For various reasons the Standard Model is typically considered as an effective description, approximating the low-energy limit of some fundamental theory in the UV. The naturalness problem of the Standard Model stems from the quadratic sensitivity of the Higgs mass to the mass scale of its possible UV completion, making it difficult to explain why the EW scale is so much lighter than the new physics scale, which, depending on the type, is pushed by the experiments to the TeV or even multi-TeV scale. One possible ingredient of the answer to this puzzle could be the Goldstone symmetry: scalar particles, such as Higgs, could be naturally light if they are associated to spontaneously broken global symmetries. Concrete and minimal realizations of this idea (e.g. \cite{Panico:2015jxa}) typically allow to push the QCD-charged new physics scale to $\sim 1$~TeV at most~\cite{Matsedonskyi:2012ym,Panico:2012uw}, where it is already partially excluded, implying enhanced fine tuning in these models. 
The Twin Higgs models are a less minimal realization of the Goldstone symmetry protection, featuring an extended symmetry structure, which allows to push the SM-charged new physics to a several-TeV scale.
While more detailed definitions of this type of models and their variations can be found in numerous papers~\cite{Chacko:2005pe,Chacko:2005vw,Chacko:2005un,Geller:2014kta,Craig:2014aea,Low:2015nqa,Barbieri:2005ri,Barbieri:2016zxn,Serra:2019omd}, we here proceed directly to the technical points relevant for the high-temperature behaviour.
We will consider the model possessing a global symmetry $SO(8)$ which is spontaneously broken to $SO(7)$ at a scale $f=0.8$~TeV, with the SM EW symmetry embedded into $SO(4)_1 \subset SO(8)$~\cite{Geller:2014kta}. Four out of seven associated Goldstone bosons are identified with the Higgs doublet. As usual for the Goldstone Higgs models, the SM interactions with fermions $q$ and gauge bosons $V$ become trigonometric functions of the the Higgs field. The masses of SM fermions derive from the following Lagrangian
\begin{equation}
{\cal L}_{\text{mass}} = 
- \frac{\lambda_q}{\sqrt 2} f \sin \frac h f \bar q q.
\end{equation}
The extended symmetry of the model includes an approximate $Z_2$ implying a presence of SM twins $\hat q$, very close copies of SM fermions charged under analogous Twin SM gauge interactions. The twin EW symmetry is embedded in $SO(4)_2 \subset SO(8)$, such that $SO(4)_1 \cap SO(4)_2=0$. The resulting twin fermions interactions with the Higgs are shifted by $\pi/2$
\begin{equation}\label{eq:qmass}
\hat{\cal L}_{\text{mass}} =  - \frac{\hat\lambda_q}{\sqrt 2} f \cos \frac h f \bar{\hat q} \hat q.
\end{equation}
At this point one can already see a sign of softened Higgs mass sensitivity to the UV physics\footnote{See e.g. Ref.~\cite{Barbieri:2015lqa} for a systematic analysis of the cancellation of the quadratic cutoff sensitivity.}. The loops of each SM and twin fermion, regulated with hard cutoffs $\Lambda_q$ and $\hat \Lambda_q$, contribute to the one-loop scalar potential at the leading order as
\begin{eqnarray}\label{eq:THdiv}
\delta V_h|_{\lambda^2}  &\simeq& \frac {\lambda_q^2} {16 \pi^2} \Lambda_q^2 f^2 \sin^2 h/f + \frac {{\hat \lambda_q}^2}{16 \pi^2} \hat \Lambda_q^2 f^2 \cos^2 h/f \nonumber \\
&=& \frac {\lambda_q^2 \Lambda_q^2- {\hat \lambda_q}^2 \hat \Lambda_q^2} {16 \pi^2} f^2 \sin^2 h/f  + const.
\end{eqnarray}
The quadratic cutoff sensitivity disappears in the limit of the exact $Z_2$ symmetry $\lambda_q = \hat \lambda_q$, $\Lambda_q = \hat \Lambda_q$, and the light Higgs mass becomes compatible with a heavy cutoff physics.

Also, as was already noticed e.g. in Ref.~\cite{Kilic:2015joa,Fujikura:2018duw}, the SM and twin states induce partly cancelling temperature corrections to the Higgs potential. To the leading order in $m/T$ expansion, we have
\begin{eqnarray}
\delta V_h(T) &\simeq& \frac{T^2}{12} (m_q^2+\hat m_q^2) 
 = \frac{T^2}{24} (\lambda_q^2 - {\hat \lambda_q}^2) f^2 \sin^2 h/f \nonumber \\
&\underset{h \ll f}{\simeq}& \frac{T^2}{24} (\lambda_q^2 - {\hat \lambda_q}^2) h^2,
\end{eqnarray}
where the last line corresponds to a thermal correction to the Higgs mass at $h=0$. 
In the following, we will simply use imbalance between $\lambda_q$ and $\hat \lambda_q$ in order to produce a negative thermal correction to the Higgs mass, leading to the high-$T$ EW symmetry breaking. Let us estimate how much of the imbalance we need. $\hat n_q$ Dirac twin fermions (each twin quark counts as 3) with $Z_2$-breaking Yukawa couplings $\hat \lambda_q > \lambda_q$ lead to
\begin{equation}
\delta m_h^2(T)_{\hat q} \simeq -\frac{T^2}{12} \hat n_q {\hat \lambda_q}^2.
\end{equation}
This correction dominates over the positive SM contribution of Eq.~(\ref{eq:SMSR}) as long as
\begin{equation}\label{eq:snr}
\hat n_q \hat \lambda_q^2 \gtrsim 5.
\end{equation}
For order-ten number of twin fermions, one is then required to have $Z_2$-breaking twin Yukawas $\hat \lambda_q$ with a size $0.1-1$. 
We would not like to introduce any imbalance between the top and the twin top sectors, as a $Z_2$ between them is a defining feature of the Twin Higgs models that keeps the masses of colored top partners well above the TeV scale at no tuning cost.
We will instead assume that $Z_2$ is broken by the Yukawas of the light twin fermions. The Yukawas of light SM fermions do not contribute sizeably to the Higgs mass in any case, while their large twin counterparts lead to (see Eq.~(\ref{eq:THdiv}))
\begin{equation}
\delta m_h^2 \simeq -  \frac{\hat n_q {\hat \lambda_q}^2} {8 \pi^2} \cos \frac {2v} f \hat \Lambda_q^2\,.
\end{equation}
Reproducing the observed Higgs mass with no fine-tuning then requires
\begin{equation}
\hat \Lambda_q \lesssim 0.6\, \text{TeV}  \sqrt{5/ \hat n_q \hat\lambda_q^2}.
\end{equation}
In the next section we will show that such a cutoff $\hat \Lambda_q$ can be related to the mass of new EW-charged fermions, which we call twin partners.
Unlike the QCD-charged top partners, such new fermions are much less constrained experimentally. With a concrete model for the cutoff physics we will also be able to quantify the amount of fine tuning associated with a non-observation of the twin partners. 

\begin{figure}[t]
\begin{center}
\includegraphics[width=220pt]{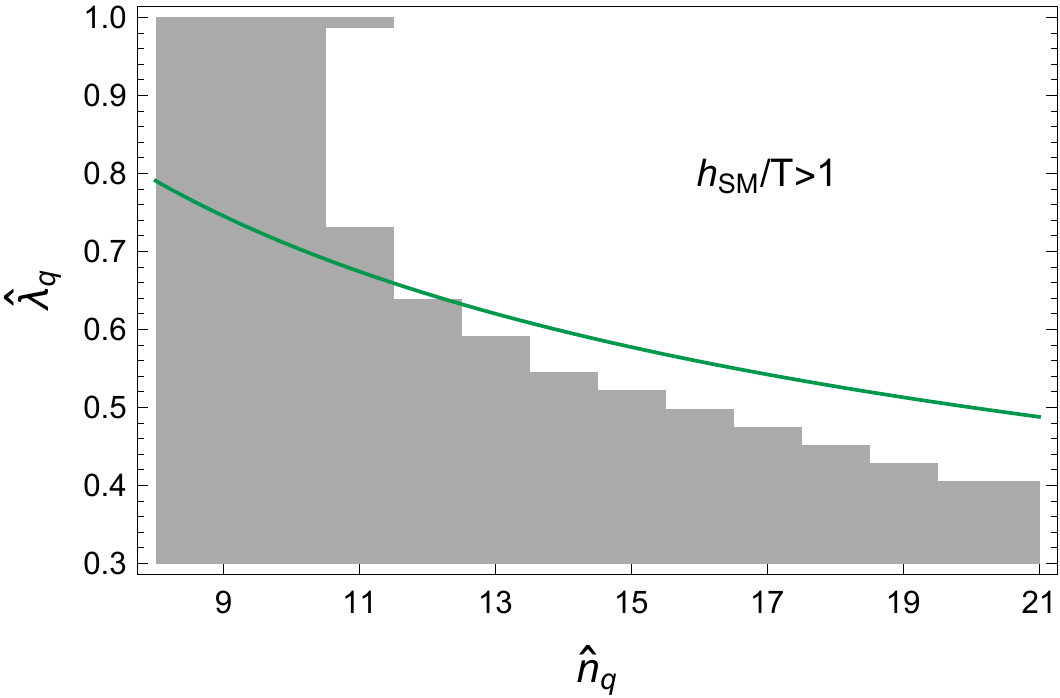}
\end{center}
\vspace{-0.32cm}
\caption{\small \it{Higgs field thermal evolution for different $\hat \lambda_q$ and $\hat n_q$. In white region the EW symmetry is broken with $h_{\text{SM}}/T>1$ for any $T<f$. In gray region $h_{\text{SM}}$ falls, at least temporarily, below $T$. Green line shows the minimal $\hat \lambda_q$ needed for SNR according to Eq.~(\ref{eq:snr}).  Maximal $\hat n_q=21$ corresponds to all twin leptons (including neutrinos which then acquire a Dirac mass) and quarks (but the top) having $Z_2$-breaking Yukawa coupling $\hat \lambda_q$.}}
\label{fig:nlambdascan}
\end{figure}

We now turn back to SNR. In Fig.~\ref{fig:nlambdascan} we show the results of a numerical computation of the Higgs field thermal evolution, for different values of $\hat \lambda_q$ and $\hat n_q$. We accounted for one-loop thermal and zero-temperature contributions from SM top quark and gauge bosons, Goldstones, and also the twin top, twin gauge bosons (we have not gauged the twin hypercharge), and the $Z_2$-breaking twins $\hat q$. 
White area corresponds to the region 
 where $h_{\text{SM}}>T$ at all temperatures within the domain of validity of our effective description\footnote{When $T / \hat m_W \propto T/ \hat g f \cos (h/f)$ becomes large, as $h$ appraoches $\pi f/2$, we expect a breakdown of perturbative expansion due to IR divergences analogously to SM. However, we do not expect that the main effect that we are investigating – SNR – will be significantly affected if we only use perturbative analysis. In particular, we know that such non-perturbative effects do not prevent EW symmetry restoration in SM, and, analogously, they are not expected to prevent the restoration of the twin EW symmetry at $h=\pi f/2$.}, $T<f$~\cite{Ahriche:2010kh,Matsedonskyi:2020mlz}\footnote{$T<f$ is required for the convergence of the Higgs loops series originating from non-linear $1/f$-suppressed Higgs interactions~\cite{Ahriche:2010kh}. Also, the convergence of fermionic loops series requires 
 $T<f \sqrt{8/\hat n_q \hat \lambda_q^2 \cos^2 h/f}$~\cite{Matsedonskyi:2020mlz}, which is only violated at large $\hat \lambda_q$ far above the minimal values needed for SNR. Such large $\hat \lambda_q$ would also lead to larger fine-tuning than what is necessary for SNR.}. We have checked explicitly that inclusion of the twin partners discussed in the next section does not change substantially the results presented in this plot.  We defined $h_{\text{SM}} \equiv f \sin h/f$ which is the quantity setting the strength of the EW symmetry breaking, so that $h_{\text{SM}}/T>1$ represents the condition preventing the baryon asymmetry washout in EW baryogenesis scenarios. In the region of light twin fermion masses $\hat m_q \propto \hat \lambda_q$ the actual boundary of the SNR region is somewhat below our estimate of Eq.~(\ref{eq:snr}). This mismatch occurs because, besides $\hat n_q$ twin fermions, also the twin top and gauge bosons contribute to SNR. Because of their large masses their contribution is somewhat suppressed and can not be simply incorporated in our previous estimate which was based on the leading term in $m/T$ expansion. 
As $\hat \lambda_q$ (and $\hat m_q$) increases, the $\hat q$ contribution to the high-temperature potential also starts deviating from the large-$T$ estimates.

\begin{figure}[t]
\begin{center}
\includegraphics[width=210pt]{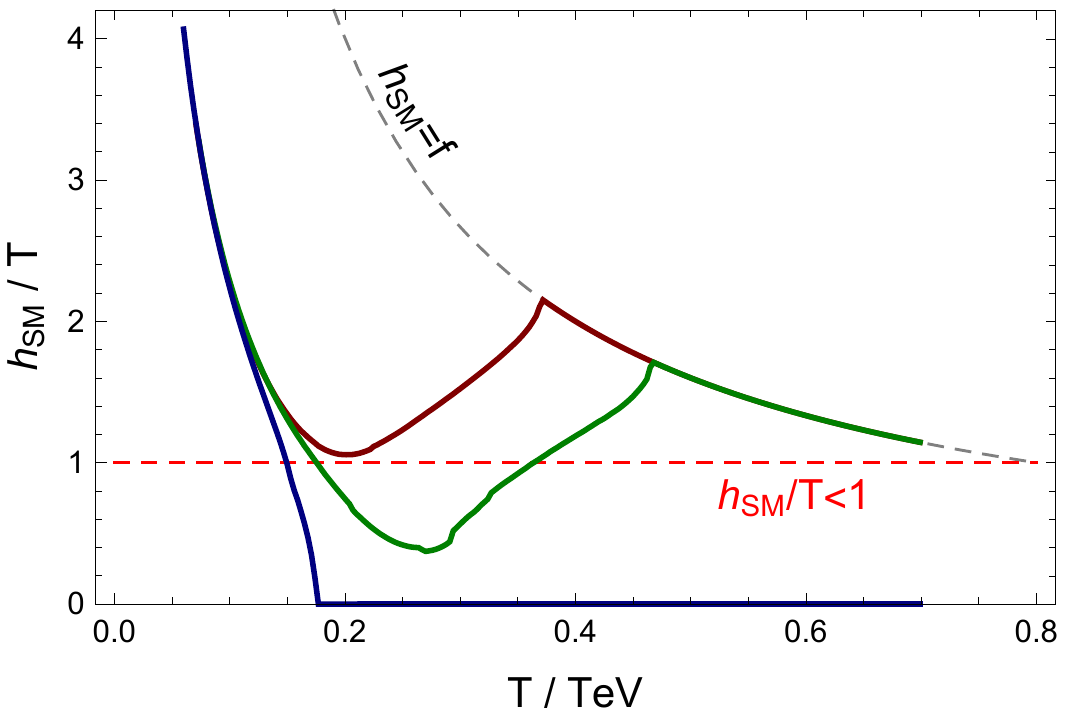}
\end{center}
\vspace{-0.2cm}
\caption{\small \it{Solid lines show the evolution of $h_{\text{SM}}/T$ in the minimum of the Higgs potential depending on the temperature, for three choices of parameters: $\hat \lambda_q=0.2, \hat n_q=9$  (blue), $\hat \lambda_q=0.5, \hat n_q=12$ (green), and $\hat \lambda_q=0.7, \hat n_q=12$ (red).}}
\label{fig:hevol}
\end{figure}

In Fig.~\ref{fig:hevol} we show three distinct types of the Higgs field evolution with temperature: EW symmetry restoration, behaviour with alternating $h_{\text{SM}}/T\lessgtr1$, and the trajectory with $h_{\text{SM}}/T>1$. In the latter two cases the Higgs field value growth with temperature is limited by the position of the minimum of the thermal potential induced by the twin fermions and gauge bosons, which is defined by $\hat m_{q,t,W}(h) = 0$, and corresponds to $h=\pi f/2$, or $h_{\text{SM}}=f$. In this minimum the EW symmetry is maximally broken, while the twin EW symmetry is restored.

We conclude that, in the proposed modification of the Twin Higgs scenario, $h_{\text{SM}}/T>1$ can be achieved across  a wide range of parameters and $T<f$. One of the most important applications can be related to the EW baryogenesis.  For concreteness, we will sketch a particular realization of the latter in the context of the Composite Twin Higgs scenario~\cite{Geller:2014kta,Low:2015nqa,Barbieri:2015lqa}, which represents one possible UV completion of the low-energy Twin Higgs model that we analysed. In this scenario the Higgs boson is a bound state of some new strong dynamics. As the universe cools down from temperatures $T\gg f$, the composite sector is initially in the deconfined phase, with no Higgs boson and no EW symmetry breaking condensate. At the temperature of the new strong sector confinement phase transition, which can be as large as $f\sim 1$~TeV, the strong sector condenses, forming in particular the Higgs boson. SNR fermions now lead to the breaking of the EW symmetry even at this high temperature. The baryon asymmetry can be generated upon this confinement phase transition, if it is of the first order~\cite{Bruggisser:2018mus,Bruggisser:2018mrt}, and remain unaffected by the sphaleron washout which is inefficient for $h_{\text{SM}}/T>1$.

\section{Modelling the cutoff physics}

Let us now analyse in more detail the cutoff physics, which is required to regulate the quadratic divergence of the Higgs potential, originating from the introduced $Z_2$ breaking.
We will promote our description of the twin SM to a two-site $SO(8)$ composite Twin Higgs model of~\cite{Barbieri:2015lqa,Contino:2017moj}, with the first site containing $\hat n_q$ elementary twin fermions, and the second site containing their composite partners. 
On the first site, we introduce elementary twin-$SU(2)_L$ doublets $\hat Q_L=(\hat u_L,\hat d_L)$, transforming as incomplete multiplets of $SO(8)$. 
We also add elementary twin-$SU(2)_L$ singlets $\hat u_R$ which are $SO(8)$ singlets too. In the following we will see that $\hat u_L$ and $\hat u_R$ acquire a Dirac mass term of the form~(\ref{eq:qmass}) and therefore they will be associated with the $\hat q_{L,R}$ fermions of the previous section. $\hat d_L$ would require $\hat d_R$ to obtain a mass, which we don't introduce for simplicity.

The second site consists of heavy twin partners, whose mass regulates the quadratic divergence of the Higgs potential. 
These partners together form complete multiplets of $SO(8)$, but their interactions and masses break it to $SO(7)$. We correspondingly split the twin partners into an $SO(7)$ and SM singlet $\psi_1$, and a {\bf 7} of $SO(7)$,  $\psi_7$.
The latter multiplet contains three SM singlets $s_i$ and two SM EW doublets, $(\psi^0_1,\psi^-)$ and $(\psi^+,\psi^0_2)$, where superscripts denote the SM electric charge. The fermions are embedded into fundamental representations of their respective $SO(8)$ as
\begin{equation}
\hat Q_L = \frac 1 {\sqrt 2}\begin{pmatrix} 0\\0\\0\\0\\ i \hat d_L\\ \hat d_L\\ i \hat u_L\\ - \hat u_L \end{pmatrix},\;\;
\psi_7+\psi_1 = \frac 1 {\sqrt 2} \begin{pmatrix} i \psi^- - i\psi^+\\ \psi^- +\psi^+\\ i \psi^0_2+ i \psi^0_1\\  \psi^0_2- \psi^0_1\\ \sqrt 2 s_1\\ \sqrt 2 s_2\\ \sqrt 2 s_3 \\ \sqrt 2 \psi_1\end{pmatrix}.
\end{equation}
 
The two sites interact via the following Lagrangian
\begin{eqnarray}\label{eq:massmix}
{\cal L} &\supset&  y_L f \bar {\hat Q}_L U (\psi_{7R}+\psi_{1R}) + y_R f \bar {\hat u}_R \psi_{1L} + \text{h.c.} \nonumber \\
& &- m_1 \bar \psi_1 \psi_1 - m_7 \bar \psi_7 \psi_7.
\end{eqnarray}
Parameters $y_L$ and $y_R$ control the mass mixing between the two sites, and $m_{1,7}$ are masses of the vector-like partners. $U$ is a Goldstone matrix which is introduced as a link between the $SO(8)$ groups of the two sites and reads in the unitary gauge
\begin{equation}
U = \left( 
\begin{matrix}
\mathbb{I}_{3} & 0 & 0 & 0 \\
0 & \cos \frac {h}{f} & 0 & \sin \frac {h}{f} \\
0 & 0 & \mathbb{I}_{3} & 0 \\
0 & -\sin \frac {h}{f} & 0 & \cos \frac {h}{f} 
\end{matrix}
\right).
\end{equation}

After mass diagonalization the approximate EW doublets mass eigenstates $(\psi^0_1,\psi^-)$ and $(\psi^+,\psi^0_2)$ retain the mass $m_7$. 
The lightest massive state $\hat u$ acquires a mass~\cite{Barbieri:2015lqa} 
\begin{equation}
\hat m_u=\frac{1}{\sqrt{2}} \frac{y_L y_R f}{\sqrt{m_1^2 + y_R^2 f^2}} f \cos \frac v f,
\end{equation}
and corresponds to the $\hat q$ fermion considered in the previous section.
Comparing this mass to the Lagrangian of Eq.~(\ref{eq:qmass}) we  identify the $\hat q$ Yukawa coupling
\begin{equation}\label{eq:lambdaq}
\hat \lambda_q \longleftrightarrow \frac{y_L y_R f}{\sqrt{m_1^2 + y_R^2 f^2}}.
\end{equation}

Using Eq.~(\ref{eq:massmix}) we can also compute the leading one-loop correction to the Higgs potential (assuming $y_{L,R}\ll1$), which is now only logarithmically-divergent
\begin{eqnarray}\label{eq:deltav}
\delta V_h &=& - \frac{\hat n_q}{16 \pi^2} \sum_{i=\psi,q} m_i^4 \log \frac{m_i^2}{\mu^2} \nonumber \\
&\underset{\log \mu^2}{\simeq}& \frac{\hat n_q}{16 \pi^2} \text{Tr} [M.M^\dagger]^2 \log {\mu^2} \nonumber \\
&=& \frac{\hat n_q}{16 \pi^2} y_L^2 f^2 \sin^2 \frac h f \left(m_7^2 - m_1^2 \right) \log {\mu^2}, \label{eq:deltaV}
\end{eqnarray}
where $M$ is the twin fermion mass matrix derived from Eq.~(\ref{eq:massmix}), which can also be found in Ref.~\cite{Contino:2017moj}.
The leading log result is enough for analytic understanding of the relevant properties, while in the numerical computations the exact expression will be used. Comparing the previous simplistic estimate of the Higgs potential in Eq.~(\ref{eq:THdiv}) with Eq.~(\ref{eq:deltav}) we see that the place of the cutoff $\hat \Lambda_q$ was taken by the masses of the twin partners, which therefore have to be relatively light to minimize the tuning of the Higgs mass.

\section{Light twin partners phenomenology}

As we will show in the next section, the naturalness arguments push the mass $m_7$ of $\hat n_q$ EW-charged twin partners $\psi^\pm,\psi^0_{1,2}$ to the TeV-scale, which makes them potentially accessible at the LHC. This represents the main phenomenological difference compared to the usual Twin Higgs models such as~\cite{Barbieri:2015lqa,Contino:2017moj}, where both EW and QCD-charged partners can (and typically are assumed to) remain beyond the LHC reach without worsening the fine-tuning.

Given a large multiplicity of the twin partners, their direct EW pair production plays the dominant role (for other production channels see~\cite{Cheng:2016uqk}).  
After being produced, the twin partners are expected to decay to SM-neutral twin fermions $\hat q$ and the singlet $\psi_1$ (if the latter are sufficiently light which we assume to be the case) and the EW gauge and Higgs bosons. Corresponding branching ratios can be estimated using the Goldstones equivalence theorem in $v=0$ approximation~\cite{DeSimone:2012fs,Cheng:2016uqk}. The interactions with the SM (would-be) Goldstones $\phi^{0}, \phi^{\pm}$ and the Higgs boson come from the mixings and the derivative interactions. The mixing part is given by
\begin{eqnarray}
y_L f \bar {\hat Q}_L U \psi_{7R} 
&\supset& y_L \bar {\hat u}_L \,\Pi.\psi_{7R}  \\ 
&\supset& y_L \bar {\hat u}_L \left(\sqrt{2}\phi^+ \psi^- - (h+i \phi^0) \psi^0_1 \right)/2, \nonumber
\end{eqnarray}
and analogously, up to signs, for the interactions with the second doublet $(\psi^+,\psi^0_2)$. $\Pi$ is a vector of Goldstone bosons.
Furthermore, the symmetry of the model admits the following derivative interaction~\cite{DeSimone:2012fs}
\begin{equation}
c_{71}\, \bar \psi_7 d_\mu \gamma^\mu \psi_1,
\end{equation}
where $d_\mu = \sqrt 2 \partial_\mu \Pi/f$ and $c_{71}$ is an order-one coefficient. After integration by parts and applying the fermionic equations of motion, this term produces interactions between the two EW doublets, the singlet $\psi_1$ and EW Goldstones, with a typical strength $c_{71} (m_7-m_1)/f$. We should also account for the fact that $\psi_{1R}$ mixes with $\hat u_R$ with the strength $\sim y_R f/m_1$.

Using this information we can qualitatively summarize the main decay channels of the EW doublet twin partners. Neglecting the kinematical factors, taking $c_{71}\sim 1$ and $m_7 \sim m_1 \equiv g_\star f$ we obtain
\begin{itemize}
\item 
$\psi^\pm$ decay to $W^\pm$ with $BR(W \hat q)/BR(W \psi_1) \sim (y_L^2+y_R^2)/g_\star^2$.
\item 
$\psi^0_{1,2}$ decay to $Z$ and $h$ with almost equal probability, and $BR(Z/h\, \hat q)/BR(Z/h \,\psi_1) \sim (y_L^2+y_R^2)/g_\star^2$.
\end{itemize}

The decay products -- $\hat q$ and $\psi_1$ -- may decay through different twin particles, and, if charged under twin QCD, hadronize. 
Following~\cite{Craig:2015pha}, we summarize the EW-neutral twin states, which can eventually be produced:
\begin{itemize}
\item
twin leptons and photons (if present): may escape the detector, unless a mixing with their SM counterparts is introduced;
\item
twin glueballs: can decay into twin photons, twin mesons, escape the detector, or produce visible displaced decays to SM via the mixing with the Higgs, depending on the size of twin $\Lambda_{\text{QCD}}$;
\item
twin mesons: can decay into SM via the Higgs mixing, twin glueballs, or twin leptons and photons;
\end{itemize}
Among the listed possibilities we here concentrate on the simplest, where the EW-neutral twin states remain undetected, manifesting themselves as a missing energy. The two-body decays of EW-charged twin partners then lead to the signatures considered in the searches for charginos and neutralinos pair production~\cite{ATLAS:2018diz,CMS:2020wxd,Aad:2019vvf,Aad:2019vnb}. We will use a very simplistic way of recasting these bounds into the bounds on our model by finding $m_7$ which would give the same number of signal events as the pairs of charginos and neutralinos with the mass equal to the experimental lower bound $m_{\text{exp}}$
\begin{equation}
\hat n_q \, {\cal L}\, \sigma_{pp\to \psi^\pm \psi^0_{1,2}}[m_7] = {\cal L}_{\text{exp}} \, \sigma_{pp\to \chi^\pm \chi^0} [m_{\text{exp}}],
\end{equation}
where ${\cal L}_{\text{exp}}$ is the integrated luminosity used to obtain the experimental bound, and we set ${\cal L}={\cal L}_{\text{exp}}$ to derive the current bounds, and use larger ${\cal L}$ to derive future projections. Multiplication by branching ratios are also performed depending on the particular decay channel. For the SUSY cross-sections we take the NLO results derived using~\cite{Fuks:2012qx,Fuks:2013vua}. We also use these cross-sections to estimate the production rate of EW partners, taking $\sigma_{pp\to \psi^\pm \psi^0_{1,2}}[m]=\sigma_{pp\to \chi^\pm \chi^0}[m]$ for wino-like $\chi^{\pm}, \chi^0$.

The bounds get weaker for higher mass of the fermions produced in the decay. The minimal $\hat m_q = \hat \lambda_q f \cos v/f /\sqrt 2$ needed for SNR can be read off Fig.~\ref{fig:nlambdascan}, and varies in the range $0.2 - 0.4$~TeV depending on $\hat n_q$. For the decay product mass $0.2$~TeV, the strongest bound comes from Ref.~\cite{CMS:2020wxd}, $m_\chi>0.73$~TeV implying $m_7>1.1$~TeV. 
On the upper side of the $\hat m_q$ mass range, the current experimental searches are not able to derive any constraint. $\hat n_q$ enhancement may lead to some exclusion, though definitely weaker than the derived above bound (because of lower sensitivity and a lower $\hat n_q$), and a more involved analysis would be needed in this case. 
 
In case if the mass splitting does not allow for a two-body decay, the three-body decays via off-shell SM gauge bosons will take place instead. In this case, the lifetime of EW-charged partners is typically longer than the twin QCD hadronization time. The bound states of produced pairs can then decay back into the SM states, significantly enhancing the experimental sensitivity~\cite{Cheng:2016uqk}. We will not consider such a compressed spectrum in the following.

Note that the bounds can also be potentially weakened if the $\hat n_q$ fermions are not exactly mass-degenerate, and a dedicated analysis would be needed for that case.

\section{Fine-Tuning}

\begin{figure}[t]
\vspace{0.1cm}
\begin{center}
\includegraphics[width=222pt]{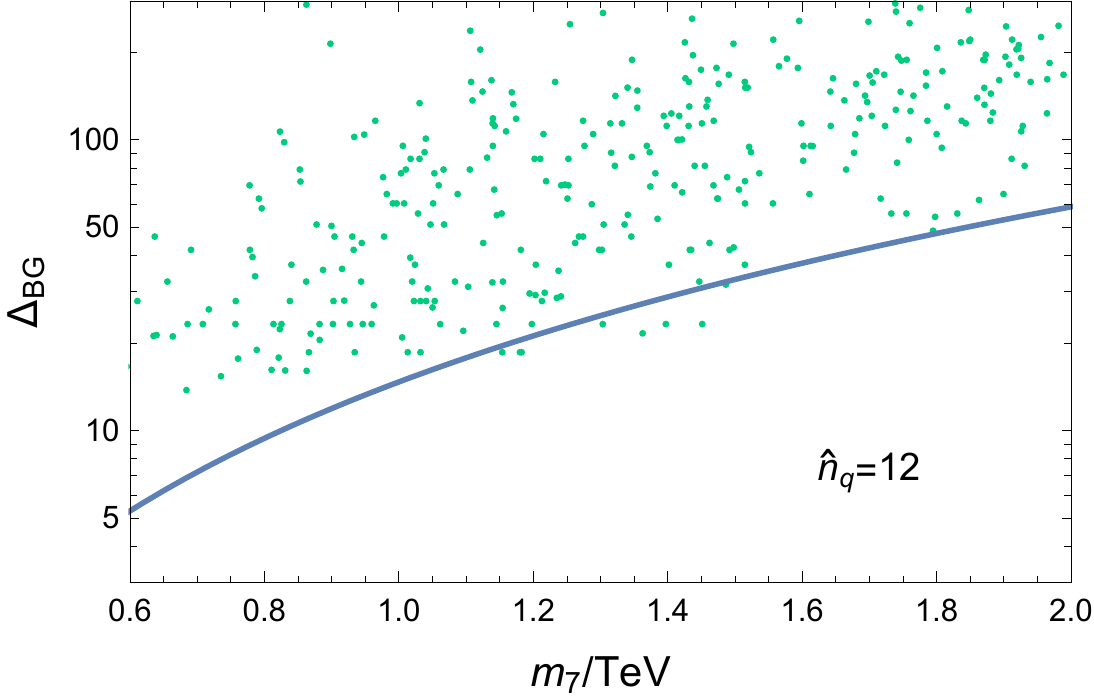}
\end{center}
\caption{\small \it{Tuning~(\ref{eq:tun}) as a function of $m_7$ for $\hat n_q=12$, for the parameter space points with SNR. Parameters are varied within the following ranges: $m_7=0.5-2$~TeV, $m_1=0.1-2$~TeV, $y_{L,R}=0.1-2$, and $\mu=3$~TeV. Blue line shows the analytic approximation of Eq.~(\ref{eq:deltaBG2}).}}
\label{fig:tunscat}
\end{figure}

To quantify the fine-tuning associated with the $Z_2$ breaking in light fermion sector we will adopt the standard tuning measure of Ref.~\cite{Barbieri:1987fn}
\begin{equation}\label{eq:tun}
\Delta_{\text{BG}} = \text{max}_{i}\left|\frac{\partial \log m_Z^2}{\partial \log x_i} \right|,
\end{equation} 
where $x_i=(y_L, y_R, m_7, m_1, \mu)$. 
Note that $\mu$ can be thought of as a physical mass of heavier particles regulating the log divergence. 
A simple analytical approximation for the minimal amount of tuning needed for SNR to happen can be derived as follows. The SM $Z$-boson mass is related to the Higgs VEV through $m_Z \propto \sin{v/f}$. The value of $\sin{v/f}$ is a result of the minimization of the Higgs potential, for which we take
\begin{equation}
V_h = \alpha \sin^2 \frac h f + \beta \sin^4 \frac h f.
\end{equation} 
Coefficients $\alpha$ and $\beta$ receive contributions from various sources, including the $Z_2$-breaking twin fermions that we introduced. If the latter give a too large correction, it has to be fine-tuned.  Experimentally,  one has to satisfy $\sin^2{v/f} = -\alpha/2\beta$ and $\alpha = - m_h^2 f^2/ 4 \cos^2 v/f$. With these expressions we can easily estimate the tuning. Given that the dominant one-loop effect of Eq.~(\ref{eq:deltaV}) is a modification of $\alpha$, we get
\begin{eqnarray}
\frac{\partial \log m_Z^2}{\partial \log x_i} & = & \frac{\partial \log \alpha}{\partial \log x_i} \\
& = & \frac{\partial}{\partial \log x_i} \frac{\hat n_q y_L^2 (m_7^2-m_1^2)\cos^2 v/f \log \mu^2}{4 \pi^2 m_h^2}, \nonumber
\end{eqnarray} 
where $m_h$ and $v$ are fixed to the experimental values.
A good estimate of the lower bound on tuning can be derived from the variation with respect to $\mu$ 
\begin{equation}\label{eq:deltaBG2}
\Delta_{\text{BG}} = \frac{\hat n_q y_L^2 m_7^2  \cos^2 v/f}{2 \pi^2 m_h^2},
\end{equation} 
where we omitted $m_1$ dependence for the latter not being directly constrained by the discussed experimental searches.  
We fix the value of $y_L$ imposing the SNR condition, corresponding approximately to $\hat n_q \hat \lambda_q^2 \gtrsim 5$ (more precisely, one should use the values of $\hat n_q, \hat \lambda_q$ which can be read off Fig.~\ref{fig:nlambdascan}),
and using Eq.~(\ref{eq:lambdaq}), so that
\begin{equation}
\hat n_q y_L^2 \simeq \hat n_q \hat \lambda_q^2 \frac{m_1^2+y_R^2 f^2}{y_R^2 f^2} \gtrsim 5,
\end{equation} 
where, again, we assumed small $m_1$. Comparison of the approximation of Eq.~(\ref{eq:deltaBG2}) with the numerical results is shown in Fig.~\ref{fig:tunscat} for $\hat n_q=12$.

Plugging the strongest current experimental bound on $m_7$ that we derived for $\hat n_q=21$, and $\hat \lambda_q \gtrsim 0.4$ (see Fig.~\ref{fig:nlambdascan}) into Eq.~(\ref{eq:deltaBG2}) we get $\Delta_{\text{BG}} \gtrsim 12$. This value marginally exceeds the typical amount of tuning obtained in the Twin Higgs models, which varies in the range $f^2/2 v^2 - f^2/v^2 = 5 - 10$~\cite{Craig:2015pha}. 

\begin{figure}[t]
\begin{center}
\includegraphics[width=220pt]{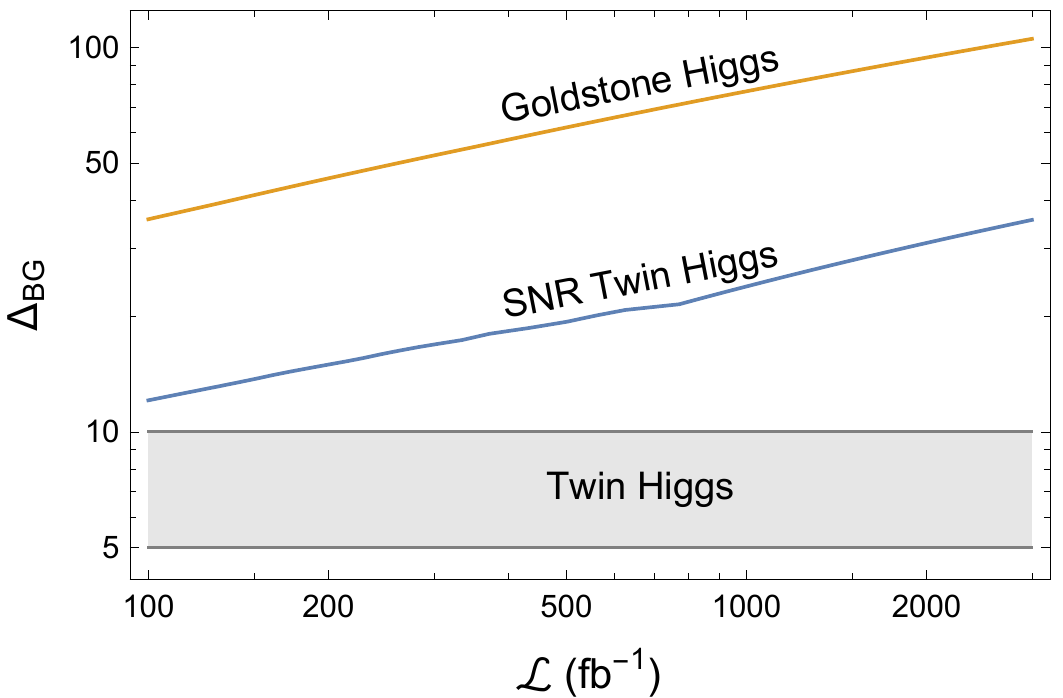}
\end{center}
\caption{\small \it{Projected amount of fine tuning in case of non-observation of top partners and twin fermion partners at 14 TeV LHC for simple Goldstone Higgs, Twin Higgs with $\hat n_q=21$ SNR twin fermions, and the vanilla Twin Higgs. Goldstone Higgs case bound can increase significantly if all top partners production channels are taken into account.}}
\label{fig:tunlum}
\end{figure}

In order to understand whether, given a worse tuning, the presence of the $Z_2$ symmetry (and hence SNR from approximate $Z_2$-copies of SM) could still be motivated by naturalness, let us compare with the case of Goldstone Higgs models without the twin symmetry protection. We will consider a Goldstone Higgs model with the symmetry breaking pattern $SO(5) \to SO(4)$~\cite{Agashe:2004rs} and SM fermions embedded in the fundamental representation of $SO(5)$. In this case, the leading contribution to the Higgs mass is proportional to the masses of QCD-charged fermionic top partners $m_*$. Corresponding tuning can be estimated using Eq.~(\ref{eq:deltaBG2}), with $\hat n_q \to N_c=3$, $y_L^2 \to \lambda_t m_*/f$, $\lambda_t^{(2\text{TeV})}\simeq 0.9$ and $m_7 \to m_*\gtrsim 1.3$~TeV, where we used the current experimental bound on the charge-5/3 top partner mass~\cite{Sirunyan:2018yun} derived using $35.9\,\text{fb}^{-1}$ of data. This results in a larger tuning $\Delta_{\text{BG}} \gtrsim 21$ for the case with no $Z_2$ symmetry. Furthermore, in Fig.~\ref{fig:tunlum} we present  the future projection of fine tuning assuming no experimental observation of the top and twin partners at the LHC. Note that we estimate the tuning in the simple Goldstone Higgs case based on the pair production of one colored partner only, while the single production and a pile up of signals from several partners are expected to significantly enhance the experimental reach~\cite{Matsedonskyi:2014mna,Matsedonskyi:2015dns,Contino:2008hi,Mrazek:2009yu}. Yet, this simple exercise shows that in case of non-observation of new physics, the twin $Z_2$ symmetry would still improve the naturalness quality, and SNR can be seen as a consequence of the latter. 

As a final comment, we would like to recall that the Goldstone Higgs models are generically tightly constrained by the electroweak precision constraints. New EW-charged fermions significantly separated from the model cutoff $\Lambda$, such as our twin partners, are known to be able to contribute to the $\hat S$ parameter~\cite{Grojean:2013qca,Contino:2015mha,Matsedonskyi:2014iha,Golden:1990ig,Barbieri:2008zt}
\begin{equation}
\Delta \hat S_\psi \approx \hat n_q \frac{g^2}{24 \pi^2} \xi (1-2 c_{71}^2) \log \frac{\Lambda^2}{m_7^2}\,,
\end{equation}
which can be helpful in resolving possible tensions resulting from the presence of irreducible contributions to $\hat S$ and $\hat T$ coming from other sources.

Our main focus in this section was on identifying the amount of fine tuning associated to the current and, potentially, future non-observation of the twin partners. 
Taking a more positive view on the subject, we can instead conclude that EW SNR (and, for instance, associated models of high-temperature EW baryogenesis) with a minimal amount of tuning motivates the existence of multiple EW charged states, which can be pair-produced at the LHC directly via EW interactions and, in the minimal case, having signatures closely resembling those of supersymmetric particles. In less minimal cases the signatures can include for instance prompt decays of pairs of EW partners into SM gauge bosons and twin states with the latter producing displaced decays into SM states. The preference for light EW-charged partners potentially accessible at the LHC represents an important phenomenological difference with respect to both the standard Twin Higgs and minimal Goldstone Higgs scenarios.

\section{Acknowledgements}

The author thanks Geraldine Servant for many useful comments on the manuscript. 
This work was supported by the Foreign Postdoctoral Fellowship Program of the Israel Academy of Sciences and Humanities.


\bibliography{biblio} 

\end{document}